\documentclass[runningheads]{llncs}
\usepackage{graphicx}
\usepackage{booktabs} 
\usepackage{enumitem}
\usepackage{cleveref}
\usepackage{color}
\usepackage{url}
\usepackage{subfig}
\usepackage{wrapfig}

\makeatletter
\newcommand*{\centerfloat}{%
  \parindent \z@
  \leftskip \z@ \@plus 1fil \@minus \textwidth
  \rightskip\leftskip
  \parfillskip \z@skip}
\makeatother

\begin{document}
\title{The Design and Implementation of a Verified File System with End-to-End Data Integrity}
%
%
\author{Daniel W. Song, Konstantinos Mamouras, Ang Chen, Nathan Dautenhahn, and Dan S. Wallach}
\authorrunning{D. Song et al.}
%
%
\institute{Rice University, Houston TX 77005, USA \\
\email{\{dwsong,mamouras,angchen,ndd,dwallach\}@rice.edu}}
%
%
%
\maketitle              
\begin{abstract}
Despite significant research and engineering efforts, many of today's
important computer systems suffer from bugs.
To increase the reliability of software systems, recent work has
applied formal verification to \textit{certify} the correctness of
such systems, with recent successes including certified file systems
and certified cryptographic protocols, albeit using quite different
proof tactics and toolchains.  Unifying these concepts, we present the
first certified file system that uses cryptographic primitives to
protect itself against tampering. Our certified file system defends
against adversaries that might wish to tamper with the raw disk. Such
an ``untrusted storage'' threat model captures the behavior of storage
devices that might silently return erroneous bits as well as
adversaries who might have limited access to a disk, perhaps while in
transit.
In this paper, we present IFSCQ, a certified cryptographic file system with strong integrity guarantees. 
IFSCQ combines and extends work on cryptographic file systems and formally certified file systems to prove that our design is correct.
It is the first certified file system that is secure against strong adversaries that can maliciously corrupt on-disk data and metadata, including attempting to roll back the disk to earlier versions of valid data.
IFSCQ achieves this by constructing a Merkle hash tree of the whole disk, and by \textit{proving} that tampered disk blocks will always be detected if they ever occur.
We demonstrate that IFSCQ runs with reasonable overhead while detecting several kinds of attacks.
\end{abstract}
\section{Introduction}
\label{sec:introduction}

Security researchers regularly find serious security flaws in critical components, even those with years of scrutiny and widespread deployment.  A promising line of recent studies have applied \textit{formal verification} techniques to significant real systems.  Such certified systems come with a machine-checkable proof that the implementation meets the specification of ``correctness''.  For example, Project Everest~\cite{everest} proved the correctness of a complete HTTPS stack by not only building and verifying the TLS implementation but also the underlying toolchain.

In this work, we examine secure file systems.  Previous verified file systems~\cite{fscq,yggdrasil,cogent} have proven their correctness with respect to important properties like guaranteed recovery from arbitrary crashes. However, none of these provide the features commonly associated with {\em cryptographic} file systems, which generally use cryptographic primitives to defend against integrity and/or confidentiality attacks, particularly under the ``evil maid'' threat model~\cite{evil_maid}, wherein an attacker has brief, physical access to the target disk, and can thus make arbitrary read and write commands to the raw storage device.

Our file system, IFSCQ, built as an extension to FSCQ~\cite{fscq}, is the first file system that not only provides proven consistency guarantees against crash failures but also proven integrity guarantees against a compromised or malicious disk. At a high level, IFSCQ defines a formal model of a malicious disk, and proves that compromise of any data or metadata will be detected via cryptographic evidence. This is achieved by constructing a traditional Merkle tree~\cite{merkle_tree} over every disk block.  The root of the tree is saved in a modeled TPM (i.e., a small trusted storage where we can keep a handful of Merkle tree root hashes), allowing us to store the bulk of the Merkle tree on the potentially vulnerable disk, while verifying its correctness every time it's re-read. 
%
As we'll see later, adding these integrity features to FSCQ required a variety of changes in FSCQ's core logic, but all our changes continued to pass FSCQ's proof obligations as well as the additional integrity properties we formalized and verified. 


We inherit the same trusting computing base (TCB) from FSCQ: we depend on the Coq proof system, Haskell compiler and runtime, and Linux FUSE (user-space file system support). Our goal was never to minimize the TCB but to verify that our cryptographic file system is ``correct'', both with respect to crash consistency and with respect to detecting disk tampering. We note that many other research projects have considered how to
verify an entire software stack~\cite{deepspec} or to generate C code with minimal runtime dependencies~\cite{cogent}. Such efforts would also be applicable to our work on IFSCQ if TCB minimization or full-stack verification was a goal.

We note that the original FSCQ work is based on the design of the Unix Version~6 filesystem; it was never designed with modern optimizations. Instead, it was meant to be simple, and to be later extended for other needs. As such, IFSCQ is similarly not intended to have competitive performance with modern production filesystems. It's meant to demonstrate the feasibility of applying formal verification toward important security properties.

\vspace{1mm} 
\noindent\textbf{Threat model.} We assume that the hardware and operating system are trusted 
but the disk can be malicious. We generalize the ``evil maid'' threat to a ``Byzantine disk,'' where adversaries can make arbitrary writes at any time. Attacks could be targeted at specific file blocks, metadata blocks, or Merkle hash blocks, or they could roll back the complete state of the disk to an previously valid state. As such, we require some sort of trusted ground truth to store and update the Merkle tree root. We borrow the term TPM (trusted platform module), included in many modern computers. We assume that our TPM provides us with a ``trusted boot'' facility, which most real TPMs support, but we diverge from real TPMs in several ways we will describe later.

\vspace{1mm} 
\noindent\textbf{Contributions.} Overall, we make the following contributions:
\begin{itemize}
  \item The design and implementation of a certified file system with cryptographic integrity
  that is proven to be both safe from crashes and to detect data tampering.
  \item A formal model of malicious storage systems, as well as desirable integrity properties 
        a file system should provide against such attacks. 
  \item A Merkle tree design, satisfying these integrity properties, which can protect
        all disk blocks (data blocks, inodes, and even unallocated blocks), while requiring
        only a small amount of trusted storage for root hashes.
  \item Evaluation results that demonstrate that our system runs with a reasonable overhead, 
  and defends against a range of malicious attacks. 
\end{itemize}

The rest of the paper is organized as follows.
Section~\ref{sec:background} describes the background needed to understand our system.
Section~\ref{sec:design} describes the design of our system.
Section~\ref{sec:proof} describes our proof engineering.
Section~\ref{sec:implementation} describes our implementation.
Section~\ref{sec:evaluation} presents our evaluation results. 
Section~\ref{sec:relatedwork} summarizes other related work. 
Section~\ref{sec:discussion} discusses possible improvements for IFSCQ, and
Section~\ref{sec:conclusion} concludes.

\section{Background}
\label{sec:background}

In this section, we provide more background on FSCQ and  file systems that provide data integrity using Merkle tree structures.

\subsection{FSCQ file system} 

FSCQ~\cite{fscq} is a formally verified file system based on Unix Version 6, with added crash consistency features using a typical logging structure. In the event of a crash during file operation, FSCQ either rolls back to the previous stable disk state, from before the file operation began, or finishes and commits the operation during crash recovery. Either way, post recovery, the disk is guaranteed to be in a consistent state. To model crashes, FSCQ introduces ``Crash Hoare logic'' (hereafter, ``CHL''), an extension of Hoare~\cite{hoare} logic pre-conditions and post-conditions, with an additional ``crash condition''. With CHL, the specification for every procedure in FSCQ includes pre, post, and crash statements with regard to the state of the file system. The developer must prove that, assuming the precondition is met before each procedure, the postcondition is met after successfully running the procedure, and the crash condition is met if there is a crash when running the procedure. FSCQ also uses separation logic~\cite{separation} to reason about disjoint disk locations.

FSCQ can be abstractly divided into two layers: ``above'' and ``below'' a ``log''. Below the log, we have a bare disk with asynchronous operations which might fail. Above the log, we have an ``abstract disk'' with reliable, synchronous block storage, including guarantees on its behavior across crashes. Most of the real file system semantics, such as files and directories, are implemented above the log layer and thus can be written without the extra code to deal with recovering after a crash. For IFSCQ, we similarly store our Merkle tree above the log to benefit from its crash consistency features; more on this below.

\begin{figure}
\begin{centering}
	\includegraphics[width=0.6\linewidth]{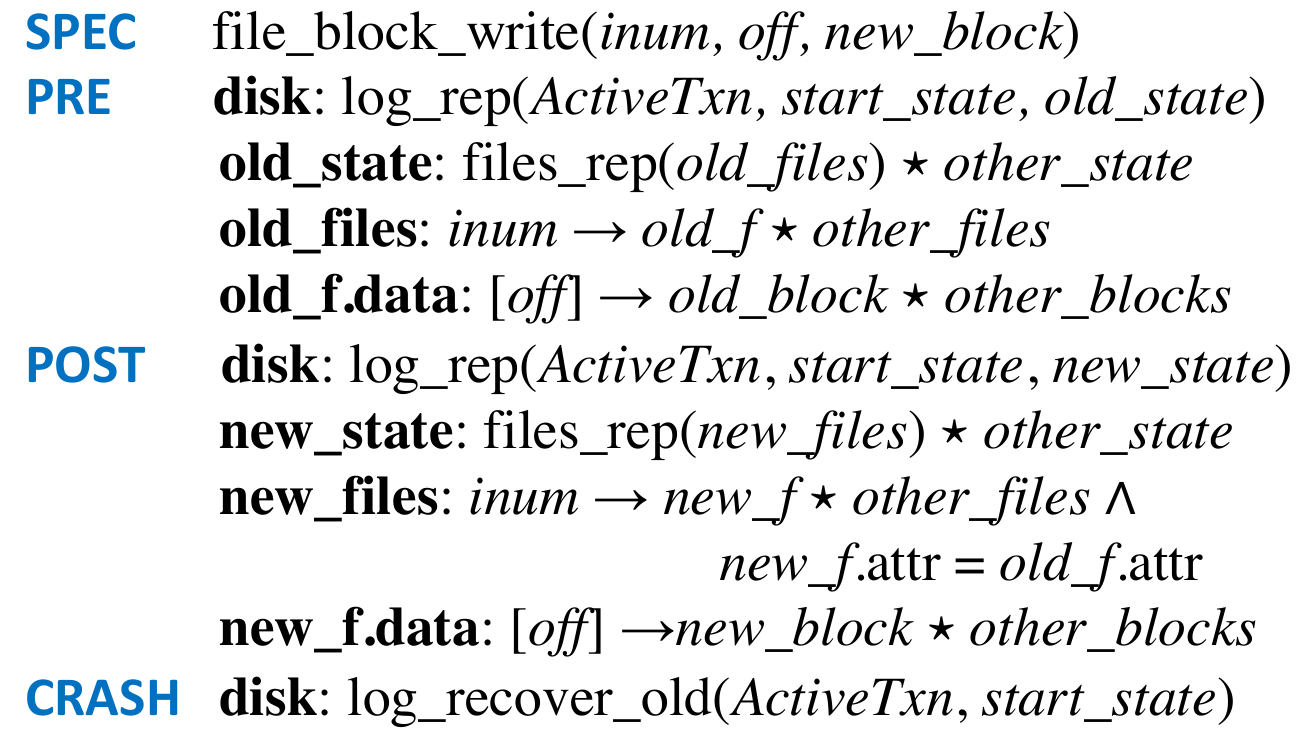}
	\caption{FSCQ file write specification\label{fig:fscq_write}}
\end{centering}
\end{figure}

To give a sense of how FSCQ is engineered, Figure~\ref{fig:fscq_write} shows the specification for a block-level write. 
The specification describes the bare disk and abstract disk using the {\em log\_rep}, with two different address spaces described in terms of files and their blocks using the {\em files\_rep}. The essence of the specification is that if the {\em files\_rep} invariant holds before the write, as described in Figure~\ref{fig:fscq_write}, it should also hold after the write. The spec also states that if there is a crash, the recovery should roll back to the pre-write state since nothing has been committed.
The $\star$ in the specification is a separating conjunction,
which allows the specification to reason about disjoint parts of the disk. 
For example, instead of representing all the files using $\wedge$ in the {\em old\_files} specification, 
$\star$ allows us to just state that {\em inum}-th file is the {\em old\_file} and the rest of the files, using $\star$ {\em other\_files},
are unchanged.

\label{sec:fscq_write}

The FSCQ file system is built in Coq, an interactive theorem prover. This implies that the 
developers should have to write the proofs as well as the implementations and the specifications. Although CHL uses various techniques such as chaining 
pre- and postconditions of subroutines in a procedure based on the control flow, the proof burden is quite high. The FSCQ authors wrote proofs that are 10$\times$ longer than their file system implementation~\cite{fscq}. After verification, FSCQ is ``extracted'' as a Haskell program, from the original Coq code. This Haskell program is then compiled by a standard Haskell compiler and integrates with the OS kernel using the standard FUSE user-level file system interface. 

\subsection{Previous cryptographic file systems}

Different cryptographic file systems provide many different properties:

\begin{enumerate}
	\item \textbf{Block-level data integrity}: Detect if there is any modification to a raw disk block. Typically done by having a checksum per block.
	\item \textbf{File-level data integrity}: Detect if there is any modification to a file. Can be achieved by checksumming or hashing the whole file, or by generating a Merkle hash tree, allowing individual byte ranges within the file to have their integrity checked without needing to read the whole file.
	\item \textbf{Metadata integrity}: file systems maintain all sorts of metadata, including names, timestamps, and permissions.
	\item \textbf{Rollback attack prevention}: It's entirely possible that an attacker could present an earlier version of a disk; detecting this requires some external, trusted metadata.
\end{enumerate}
	
A common tactic to protect integrity is to use a Merkle tree structure~\cite{merkle_tree}. A full discussion of the properties of Merkle trees is beyond the scope of this paper, but an important property to note is that Merkle trees require $O(\log n)$ reads to both validate a disk read or update themselves after a disk write. 

\begin{table*}[t!]
\centerfloat
\scalebox{0.8} {
{\small
    \begin{tabular}{| l | c | c | c | c | c |c | c| c| c| c| c|}
    \hline
         & Tripwire~\cite{tripwire} & IFS~\cite{ifs} & Iris~\cite{iris} &  SiRiUs~\cite{sirius} & IPFS~\cite{ipfs} & GFS~\cite{gfs} & ZFS~\cite{zfs} & Btrfs~\cite{btrfs} \\ \hline
        Block-level data integrity  & &  & $\circ$ & & $\circ$ &$\circ$ & $\circ$ &  $\circ$ \\
        File-level data integrity  & $\triangle$ & $\circ$ & $\circ$& $\circ$& &$\circ$ & $\circ$& $\circ$ \\
        Metadata integrity  & &$\circ$  & $\circ$& $\circ$& $\circ$ &$\circ$ &$\circ$ & $\circ$\\
        Runtime integrity checks  & &$\circ$  & $\circ$&$\circ$ & & $\circ$ &$\circ$ & $\circ$ \\
        Rollback prevention  & & &$\circ$ & & & & & \\  \hline
    \end{tabular}
}
}
        \vspace{2mm}
	\caption{Cryptographic features. $\circ$ means that it provides the feature. $\triangle$ means that the guarantee only holds on selected files.}
	\label{tab:features}	
\end{table*}

A notable example is ZFS~\cite{zfs}, which includes cryptographic checksums alongside every disk address pointer; each data block is then compared against the checksum for every read. Every address pointer is itself also saved in a data block, thus the pointer to that data block will enable that block's integrity to be checked when read. This continues all the way to the file system root. Table~\ref{tab:features} describes many other such file systems. However, unlike IFSCQ, none of these previous cryptographic file systems have been formally verified.

\section{Design: How to add a Merkle tree}
\label{sec:space}
Our threat model is a trusted computer
with a trusted operating system but an untrusted disk. Variations
on this threat model are common in the persistent authenticated
data structures (PADS) literature (see, e.g.,~\cite{pad,auth_dict}). An untrusted
disk can modify any data or metadata block at any time;
it can also launch rollback attacks on the whole disk,
providing correct (though outdated) hashes. 

Our goal is to minimally extend and integrate integrity checks into
the FSCQ proof system and file system implementation.
Doing this requires answering two design questions: 1)
what approach to use to provide data integrity, 2) at what
layer should we add this property into the file system.


Our first question is answered by using a Merkle
tree~\cite{merkle_tree}, even though PADS have been built
with a variety of other data structures. Unlike PADS, which
might grow and shrink over time, or might support queries
against older ``versions'' of themselves, we know the size
of the disk at the start, and it's not going to change. Similarly,
we only need to validate the ``current'' version of the file system,
since FSCQ has no support for historical snapshots.

To answer the second challenge, we compared several possible designs:
%

\vspace{1mm} \noindent\textbf{Above the file system.}
A seemingly obvious design is to build an integrity layer that can run
above {\em any} file system. This could be done for individual files, with
custom libraries used to read and write to those files, and with
the Merkle tree metadata written into separate, auxiliary files.
This roughly describes how BitTorrent's {\tt .torrent} files
are designed. 

Another alternative would be to implement an entire file
system-in-a-file, relying on FSCQ to provide us with a single, large,
flat file. We wouldn't have to worry about crash recovery, but we
would have to reimplement all the core abstractions of a file system:
files, directories, and so forth. Furthermore, FSCQ's FUSE interface
does not expose the internal transactional log system, without which we
would not be able to keep the Merkle tree synchronized with the
associated file data across potential crash events. We rejected this
design because the proof and implementation burden would amount to
reimplementing most of FSCQ from scratch.

\vspace{1mm} \noindent\textbf{HMAC below the file system.} 
Another seemingly obvious design is to build an integrity layer {\em
  under} FSCQ, replacing its raw block interface with calls to our new
code. The challenge with this design is that it forces the integrity
layer to deal with the crash-consistency issues that FSCQ normally
hides from the layers above it. This would add a huge amount of
complexity to the integrity system, since it would need to deal
with unreliable disks.

If we pressed on, we might
maintain a keyed hash (HMAC) of every block, without any sort of
Merkle tree. The HMAC key could be stored in a TPM. Every
block read would have a corresponding hash read and
verification. Every block write would have a corresponding hash write.
An evil disk would not know the key, so could not forge the
hashes.

This design is closely related to how Apple's
FileVault~\cite{filevault} and Microsoft's BitLocker~\cite{bitlocker}
are implemented, except those systems are generally focused on 
confidentiality concerns, where our goal is to provide integrity
guarantees.

Most obviously, this HMAC design lacks the ability to defeat rollback
attacks, since there are no semantics to indicate the age or version
number of a block in relation to its peers. There's no ``root hash''
to check against. Furthermore, this design, running below FSCQ's log
layer, would create significant proof burdens by virtue of running
directly on the unreliable, raw disk. Consequently,
we rejected this design.


\vspace{1mm} 
\noindent\textbf{Inside of FSCQ.} 
\label{sec:space_crash}
At this point, the only feasible designs left require adding the
integrity system directly into FSCQ itself. This would allow us to
benefit from FSCQ's crash consistency protections, but it would also
mean that any changes we make to FSCQ could require us to potentially
redo some or all of FSCQ's proofs---a significant engineering burden.
Furthermore, we must grapple with exactly where in FSCQ the integrity
layer should go.

IFSCQ should inherit the crash consistency guarantees provided by FSCQ. 
Without crash consistency, we could have a mismatch of
hashes after a crash even in the absence of attacks! The transactional
log structure of FSCQ guarantees that either all of the updates get
applied or none do, so it's just as essential to the correct operation of
IFSCQ. Of course, FSCQ was never engineered to operate
above a malicious storage system. The implications of this threat
model had design impacts throughout our extensions to FSCQ.



\section{Design: Integrating with FSCQ}
\label{sec:design}

In this section, we describe how we built our IFSCQ integrity layer in FSCQ.

\subsection{Implementation design}\label{sec:impl_design}

\vspace{1mm} 

\noindent\textbf{Building a Merkle tree.} In Section~\ref{sec:space}, we described our reasoning behind building a Merkle tree directly into the verified file system. We now describe how we build our Merkle tree inside FSCQ. The seemingly obvious approach is to leverage the existing tree structure from the directory tree. File systems such as IPFS~\cite{ipfs} and ZFS~\cite{zfs} use fat pointers, where every block address also has a cryptographic checksum of the data block about to be read. This class of design integrates in a straightforward way with existing file systems, allowing for verification from the root of the file system all the way to the individual files' disk blocks. Furthermore, the whole system can be implemented without adding any additional disk read or write operations, although fewer block pointers can be included in a given size of disk block. This seems like a straightforward approach to retrofitting integrity checking features into an existing file system without creating undesirable performance issues, although it's worth noting that any block write will potentially need to generate updates to every block that points to it, and all of these writes will require updates all the way to the root block. This performance issue would be exacerbated by potentially deep trees of file system directories.

An alternative approach would be to build the Merkle tree around the notion of a flat range of blocks, without any knowledge of files or directories, nor any concept of free or allocated disk space. Similar to shadow memory \cite{shadow_mem}, we can have shadow blocks, each containing an array of block checksums. We then can construct a Merkle tree over these shadow blocks. This yields a ``fat'' tree~\cite{fat_tree}, with a very wide fan-out, meaning that the depth is much more shallow than we might get if we followed the directory tree. With a terabyte filesystem, we only need a Merkle tree of depth 4. So, every regular disk write would require four additional writes to the Merkle tree. 
An additional benefit of this approach is that it requires minimal changes to most of the logic and proofs used by FSCQ. Given the high engineering burden associated with making changes and fixing proofs, this sort of orthogonality of design is a huge benefit.

A real-world concern arises with uneven write pressure on the disk: The Merkle tree's root block is rewritten for {\em every} disk write! While a full discussion of disk hardware limitations is beyond the scope of this paper, we note that all modern SSDs maintain a virtual block mapping, allowing them to remap hot disk blocks, and thus level out storage wear across the device. As such, our design certainly accelerates disk wear on writes, but it should not be any worse than an equivalent number of writes to random disk locations.

Another important design consideration is how the Merkle tree should interact with the crash recovery system. We don't want the Merkle tree to lose synchronization with the data it is protecting. So long as the Merkle tree is ``above the log layer'' and thus benefits from FSCQ's existing crash recovery logic, we should be able to rely on FSCQ to maintain this synchronization, in the case of a benign disk.

\vspace{1mm} 

\noindent\textbf{Defense against rollback attacks: TPM.}\label{sec:roll_back_tpm} We store the Merkle tree in the same disk where we save the data. This implies that it is vulnerable to rollback attacks---a malicious disk can easily roll back both the Merkle tree blocks and the regular disk blocks in a consistent fashion. In order to prevent this type of attack, we save the root hash of the Merkle tree in a modeled trusted storage system with limited, persistent state. This allows us to transfer trust from the TPM-sourced root to the untrusted, disk-sourced leaves of the Merkle tree and file system blocks we are reading.

We cannot completely avoid the crash recovery mechanism, even when running our Merkle tree completely above the log layer, because FSCQ holds out the possibility that crash recovery might roll the file system back to the previous good state. FSCQ would also then roll back our Merkle tree, but it wouldn't roll back the TPM's copy of the Merkle tree root. Consequently, our TPM must hold {\em two} roots: the last stable root, corresponding to the safely committed disk state of the file system, and a potentially unstable root, corresponding to file system writes that are not yet committed. 

This design means that we are conceding a specific rollback attack: the malicious disk can simulate a crash and force a recovery to the previous stable state. Of course, this is indistinguishable from an actual disk crash, which is already something that FSCQ is engineered to support, as best it can, so IFSCQ can do no better than this. (In modern, high performance file systems, there may be a substantial volume of disk writes buffered in memory, possibly written out-of-order to the disk.) 

Obviously, our modeled TPM is different from the TPM devised by the Trusted Computing Group. Our modeled TPM simply stores and retrieves hashes whereas the real TPM does many other things, including managing cryptographic keys and monitoring the computer's boot process. We note that a real TPM experiences very little write pressure during regular operation of a computer, but our modeled TPM is updated on every disk write. As discussed above, we now have to worry about wearing out the Flash storage inside the TPM. How might we address this with real-world TPMs hardware? We note that Flash memory is exceptionally cheap. Adding a dedicated Flash chip to the TPM would have only modest power and cost impact to a modern laptop computer, and this would allow the creation of a ring buffer of old root hashes, rather than just keeping the last two. On a filesystem supporting versioning or snapshots, this would be essential. (Our Haskell implementation of our modeled TPM simply stores the root hashes on a different disk.)

\subsection{Proof design}\label{sec:proof_design}
Previous certified file systems assume that disks may experience benign failures, but not malicious attacks. We need to model these behaviors and design our system and its proofs around them.

\vspace{1mm} 

\noindent\textbf{Modeling an evil disk.} In the real world, data integrity violations might be caused by hardware errors, software bugs, or malware~\cite{ensure_data_integrity}. Studies have shown that disks will occasionally flip bits~\cite{hw_data_corruption}, file system bugs can occasionally corrupt disks~\cite{explode}, and ransomware might even encrypt a disk's data. Although data integrity violations can result from any of these causes, the problem is fundamentally the same: when reading a block we expect to read the last value written at that block. 

Previous FSCQ work used an execution model where it describes how it should behave on reads and writes to the disk. The disk itself is a simple array of blocks and the file system performs reads and writes to these blocks. For IFSCQ, we first need to model the way an evil disks behaves. One way of modeling an evil disk is, just like malware or hardware errors, the evil disk can make an arbitrary write to any arbitrary block address at any time. We can model this in a proof system by tracking the ``true'' state of the disk alongside a list of ``tampered'' blocks, allowing us to make and verify assertions about what should happen if the set of blocks being read for any given operation includes a tampered block.

An alternative strategy would be to have a ``flaky'' disk, where a block read operation might return the legitimate value or might return something else. We then would need to prove that any operation that was given a bad input would detect this when it happens and halt the computation. This leads to an inductive proof of the file system's integrity, where every read operation is verified as it happens, and the file system halts if a verification operation fails. Consequently, a running file system implies that every prior read was correctly verified.

The second strategy avoids the need for proofs to reason about a list of malicious writes to blocks that may well never be read. Instead, the proof engineering is much simpler if we verify every disk block as it is read.

\vspace{1mm} 
\noindent\textbf{Formalization of data integrity.} 
With our malicious disk model, we require that any faulty block read will result in the file system halting. To prove this, we'll use a strategy where every file system operation will maintain two lists of disk blocks: the blocks we expect we will read, and their correct values, along with the blocks we have already read, including their possibly incorrect values. We must then prove assertions based on these block lists: if anything was tampered when read, the file system will halt.

In a ``real'' implementation, tracking all of this state requires extensive storage, have significant performance impacts, and is not easy, since by definition we don't know which block reads are good or bad until we check the hashes. In Coq, we can maintain these lists as {\em ghost variables}, representing state that we wish to use for our proofs, but which should not be included in the extracted Haskell code that actually runs the real file system. We can use these ghost variables to create a proof burden for IFSCQ to show it handles tampered blocks without needing to maintain this state at runtime.
We can also replace real hash functions with symbolic functions that are invertible. This allows our proofs to assume that hash equality implies block equality, and thus we've ``proven'' that we read the correct block, even though this is not true with real hash functions where an attacker could ostensibly find a hash collision.

The core fundamental contribution of this work is that we reason about integrity failures of the system. We proved that our file system will halt once it detects any integrity failure during operations. Thus, we add integrity failure condition to the proof system. As CHL extends traditional Hoare logic with a ``crash condition'', describing what happens in the event of a crash, we needed to add an {\bf integrity failure condition} so that we can distinguish a crash and a failure. 
If the file system halted, but the block lists are equal to one another, the crash condition should hold true. However, if the block lists don't match, then the integrity failure condition must hold true. Figure~\ref{fig:slog_code} shows a pseudocode of the {\em safe\_write} that writes to a data block and updates the Merkle tree accordingly. The integrity failure condition \texttt{IFAIL} states that if there is any integrity failure during this operation, the disk will revert back to the initial state since nothing has been committed yet.

\section{Proof Engineering}
\label{sec:proof}

This section describes in detail the properties we have proven in our system.




\begin{figure}
\centerfloat
	\subfloat[Safe Write Specification\label{fig:safe_write_spec}]{{\includegraphics[width=7cm]{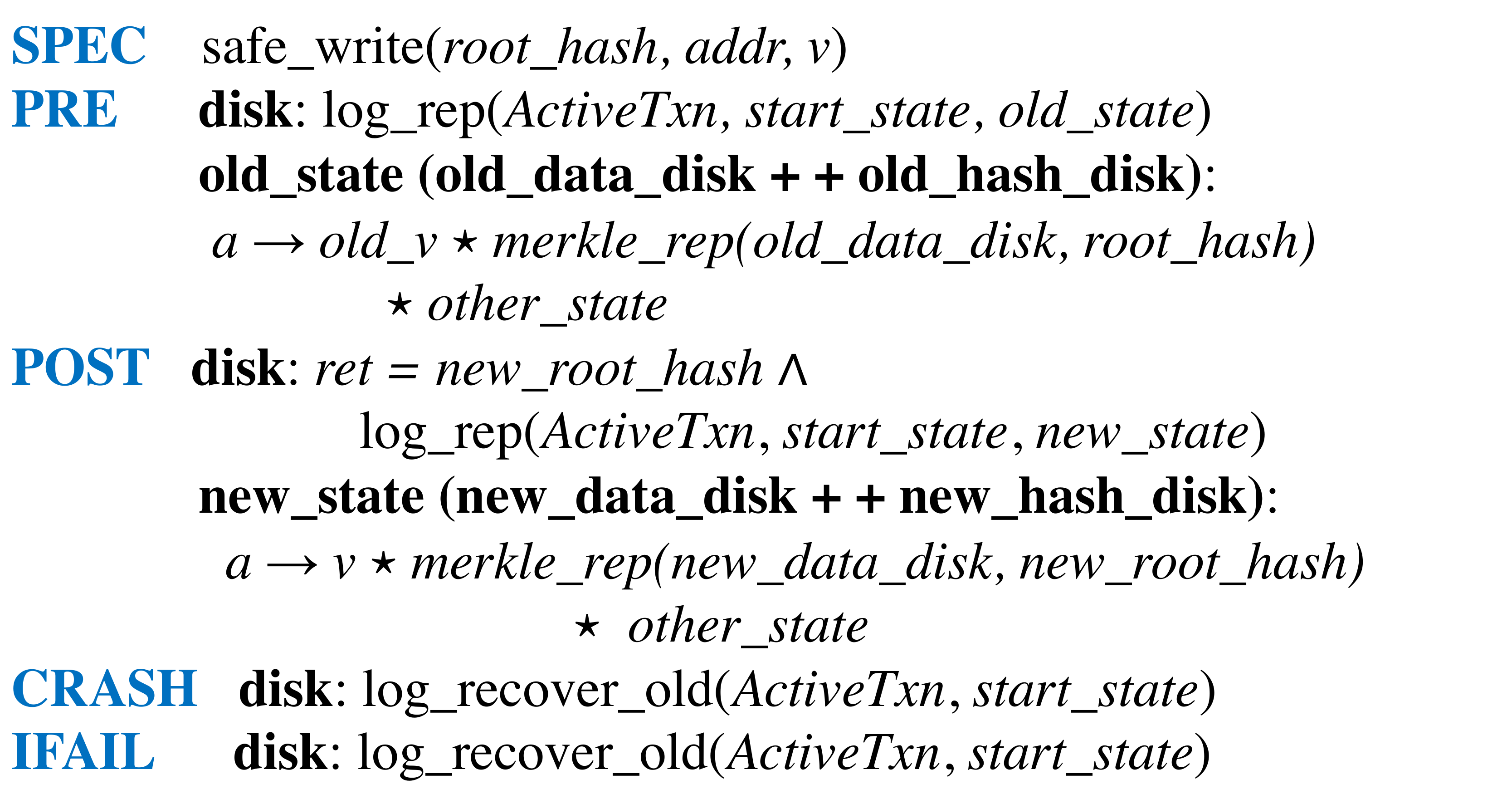} }}%
    \subfloat[Read Specification\label{fig:read_fail}]{{\includegraphics[width=7cm]{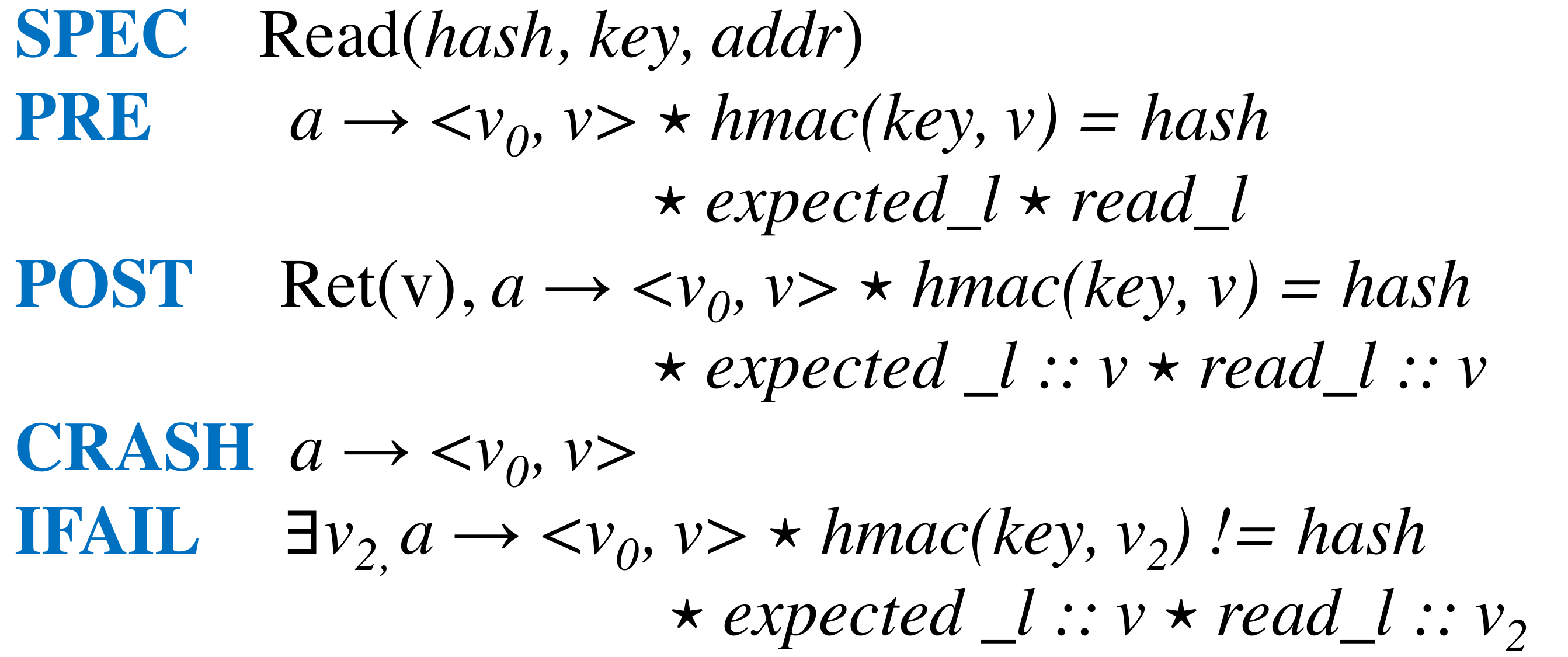} }}%
    \caption{Specifications of Safe Write and Read}%
    \label{fig:example}%
\end{figure}

\vspace{1mm} 
\noindent\textbf{Merkle tree invariant.} We prove that our system provides data integrity by proposing a {\em Merkle tree invariant} and showing that it holds for every top-level procedure. For example, the specification of the \texttt{safe\_write} uses \texttt{merkle\_rep} which is the Merkle tree invariant as shown in Figure~\ref{fig:safe_write_spec}. The \texttt{merkle\_rep} states that the leaf nodes of the Merkle tree match with the hashes of the data blocks from the disk, and hashes of the children should match with their parents. We then prove that the Merkle tree invariant is preserved for every higher-level procedure calling SLog's \texttt{safe\_read} and \texttt{safe\_write}. The \texttt{safe\_read} does not update anything so the proof is trivial. However, for \texttt{safe\_write}, we must prove that the Merkle tree invariant still holds after the write to the disk.

The \texttt{safe\_write} procedure described in Figure~\ref{fig:slog_code}, which has a specification described in Figure~\ref{fig:safe_write_spec}, writes a value \texttt{v} to a block with address \texttt{a} and also updates the Merkle tree. The \texttt{start\_state} in the precondition is the bare disk, and the \texttt{old\_state} is the newly generated disk if updates on the log were to be committed on this bare disk. The \texttt{old\_state} get updated during the operation of \texttt{safe\_write} and becomes the \texttt{new\_state} in the postcondition. The \texttt{old\_state} states that address \texttt{a} has value \texttt{old\_v} and has a Merkle tree. The Merkle tree uses data part of the \texttt{old\_state} which is \texttt{old\_data\_disk} to generate the tree and has the root hash \texttt{root\_hash}. Then we have to prove that the Merkle tree invariant still holds for the \texttt{new\_state}.


As shown in Figure~\ref{fig:slog_code}, data block updates happen in \texttt{Log.write\_data} and Merkle tree updates happen in \texttt{MERKLE.update\_hash}. Therefore, we only prove in \texttt{MERKLE.update\_hash} that the hashes of the children nodes match with the parents. The way we have implemented the Merkle tree allows us to simplify proving this Merkle tree invariant since we only update the nodes on one path at a time. The other part of the tree is still the same with the tree before the \texttt{MERKLE.update\_hash}, and thus the Merkle tree invariant still holds for the rest of the tree. We have proven that updated nodes satisfy the parent-children Merkle tree relationship and thus, it holds for the whole updated tree. 

For \texttt{SLog.safe\_write}, we prove that leaves of the Merkle tree match with the hashes of the regular data blocks. Since \texttt{SLog.safe\_write} only changes one data block, we only have to prove that the updated block's hash value match with the corresponding leaf of the Merkle tree.



\vspace{1mm} 
\noindent\textbf{File system only accepts good blocks.} As discussed in Section~\ref{sec:proof_design}, we have proven that two lists of blocks, one for the blocks we expect to read and one for the blocks we have actually read, are a match for every procedure in IFSCQ or it will halt. Instead of explicitly stating that two lists should match for every procedure, we have extended CHL so that any procedure that uses the CHL proof system should verify the blocks that are being read. These two lists are \textit{ghost variables} and do not show up during execution. We also reason about integrity failures distinguishing failures from crashes by comparing the two lists.

\begin{figure}
	\center
	\includegraphics[width=0.7\linewidth]{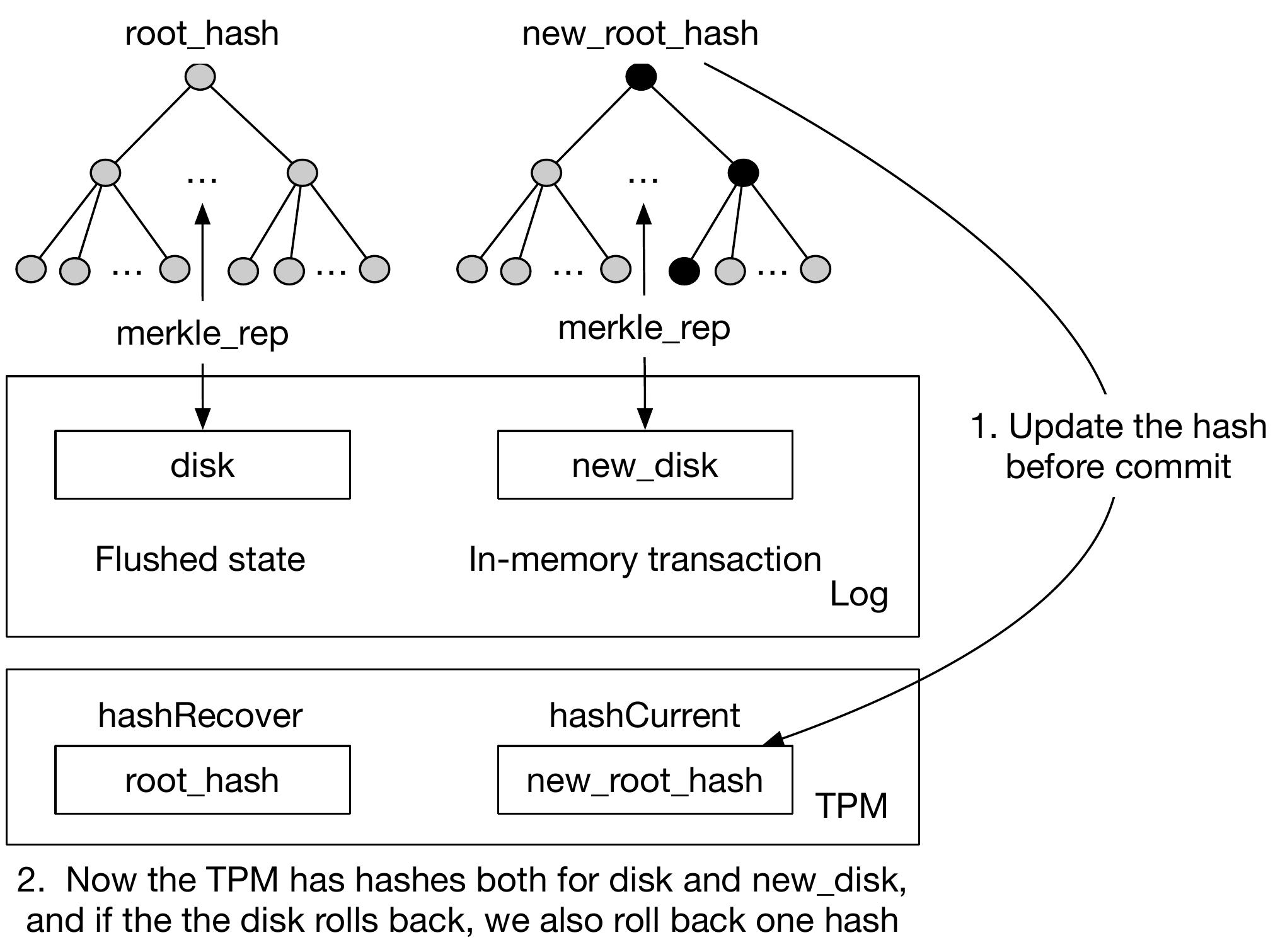}
	\caption{Merkle tree abstraction.\label{fig:merkle_tree}}
\end{figure}

 \vspace{1mm} 
 \noindent\textbf{TPM.} We have implemented and proved that if the top-level procedures complete without failure, the TPM either updates the \texttt{hashCurrent} with the new hash for the file writes or it stays the same for the file reads, as shown in Figure~\ref{fig:merkle_tree}. Also, our TPM holds two most recent hashes, \texttt{hashCurrent} and \texttt{hashRecover}, so that it can handle two possible outcomes of a crash and recovery.

\vspace{1mm} 
\noindent\textbf{Modeling hashes.} Our system relies on hashing to ensure integrity, so we need to model a hash function. One challenge is how to model collisions. DFSCQ\cite{dfscq} uses checksums to ensure integrity of the log. DFSCQ's model has a set that tracks of all the inputs that have been hashed; if a new input yields a previous output, the model enters an infinite loop. For IFSCQ, this model is inadequate for our Merkle updates. Instead, to lessen our proof burden, we modeled a perfect, symbolic hash function, without any possibility for collisions. During execution, we substitute a Haskell implementation of HMAC-SHA256.

\vspace{1mm} 
\noindent\textbf{Integrity failure condition.} As we show in Figure~\ref{fig:read_fail}, we have added an ``IFAIL'' state as one of the post-conditions to \texttt{Read}, corresponding to what will occur if there's a hash integrity failure. The pre-condition states that the most recent value that has been written to the disk address \texttt{a} is \texttt{v} and that its hash value matches with \texttt{hash}. We also have two lists of blocks, one for the blocks we expect to read, \texttt{expected\_l}, and one for the blocks we have actually read, \texttt{read\_l}. The post-condition states that it should return \texttt{v} successfully and \texttt{expected\_l} and \texttt{read\_l} both should have  \texttt{v} added to the end since it is both the expected and actually read value. The integrity failure post-condition states that some value \texttt{v2} was read from the disk and its hash didn't match, thus \texttt{read\_l} should have \texttt{v2} at the end and \texttt{expected\_l} should have \texttt{v} at the end breaking the data integrity property. Another example is shown in Figure~\ref{fig:safe_write_spec}, where the integrity failure post-condition of the \texttt{safe\_write}, as in most, is same as the disk crash post-condition. Where integrity failure cannot simply happen, such as \texttt{Write} primitive operation, we state \texttt{False} which is the strongest argument we can make. The addition of this new condition has created extra proof obligations for every procedure in IFSCQ but the proof burden was not high as it was mostly repeat of proofs for the crash conditions.

\vspace{1mm} 
\noindent\textbf{Putting it together.} Altogether, we have implemented and proved that all the blocks we read during normal file operations are good and if not, the file system will halt. Our proof also shows that the Merkle tree invariant will hold for every top-level API. Moreover, we proved that the TPM will always have the correct root hash of the disk. Finally, it is important to note that we have maintained all the existing FSCQ proofs. Therefore, we can state with a high degree of assurance that IFSCQ has all the same crash consistency guarantees as FSCQ in addition to all of its integrity guarantees.

\vspace{1mm} 
\noindent\textbf{What is proven, implemented, and extracted.} Our proofs of IFSCQ's properties are based on several unproven parts in the system; FSCQ does much the same. Our hash function, TPM, and execution model are all idealized models, allowing us to produce proofs about other parts of the system. These models are not extracted to Haskell code; we instead use separate Haskell implementation of those components during runtime. We do not provide proofs that Haskell implementation follows the Coq models.

Our Merkle tree and the interaction with the TPM are implemented and proven using Coq. Along with the models, implementations, and proofs, we prove that if the evil disk returns a bad block, the file system will halt. Finally, the Coq implementation will be extracted and compiled with the Haskell implementation of models along with the Haskell FUSE library to run a user-level file system.



\section{Implementation}
\label{sec:implementation}

\begin{figure}[t!]
	\center
	\includegraphics[width=0.5\linewidth]{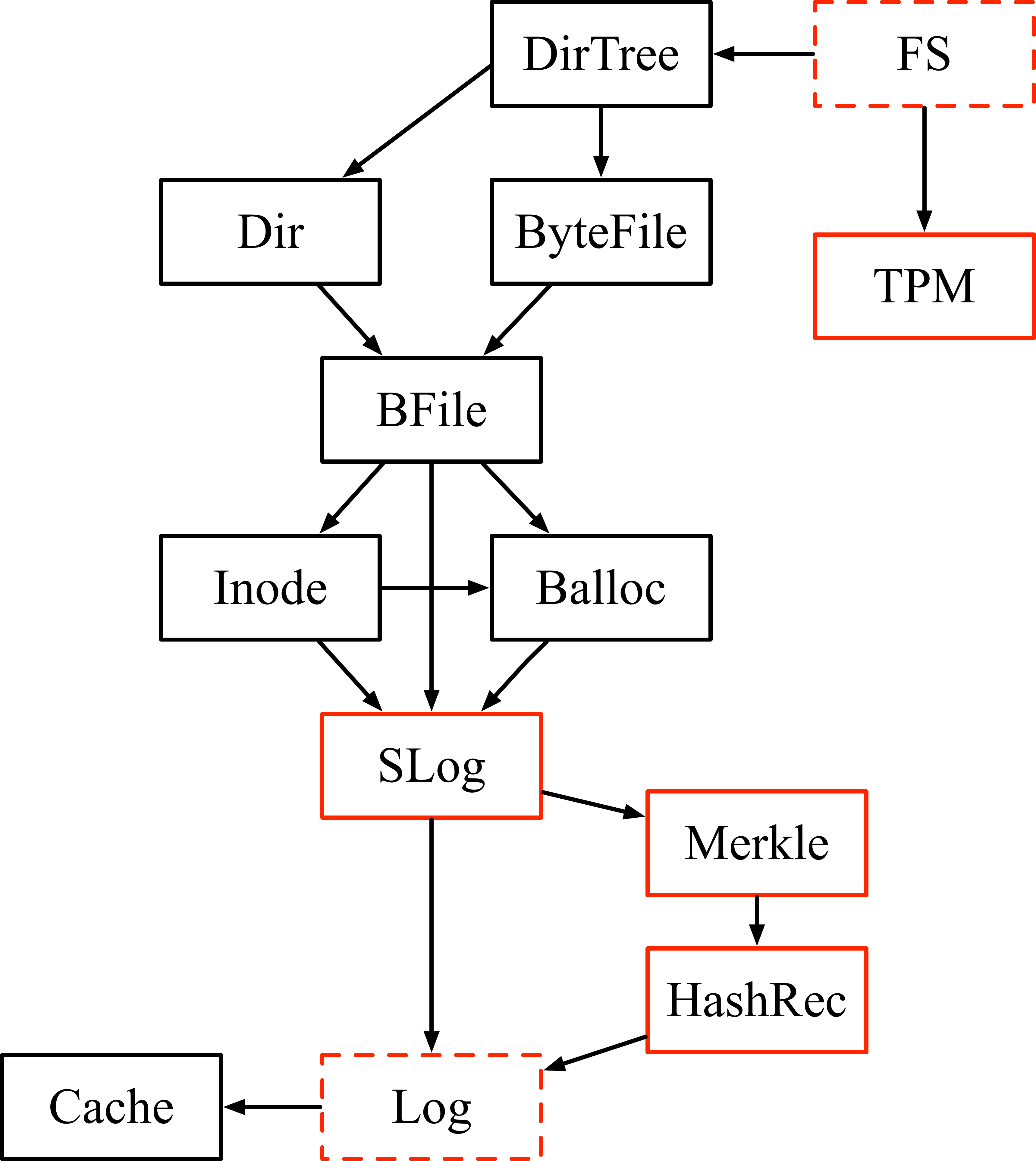}
	\caption{Components of the IFSCQ system. A red box indicates a new component, relative to FSCQ. A red dotted box indicates a FSCQ component with significant changes, and a black box indicates an FSCQ component with minimal changes.\label{fig:system_dia}	}
\end{figure}

This section describes in detail how we have implemented the system, using the Coq proof assistant.

\subsection{System overview}

IFSCQ is built on the original FSCQ, reusing many of its existing components with minimal changes.
Figure \ref{fig:system_dia} shows the IFSCQ architecture. The FS layer implements the system calls that connect FUSE and the file system. The FS calls the directory tree layer, DirTree, which manages directories and files. Dir implements the abstraction for directories and BFile implements a block-level interface for files. The TPM module models the storage of the root hash of the Merkle tree. SLog is a wrapper for the Log which does the appropriate Merkle tree operations for each corresponding disk read and write operation. The Merkle layer works alongside HashRec which translates it to disk blocks. 

As seen in Figure \ref{fig:system_dia}, the bulk of our efforts occurred in two areas: creating 
a high-level file system abstraction that's aware of our TPM and Merkle tree system, as well as a low-level extension to FSCQ's Log system, which manages writing state to disk and recovering from a crash.  We discuss these efforts in additional detail below.

\vspace{1mm} 
\noindent\textbf{TPM and FS.} The FS layer is our top-level abstraction for managing file system activity, including calls to deal with hash generation and verification, as well as updating the trusted state in the TPM.


\begin{figure}
	\center
	\includegraphics[width=0.7\linewidth]{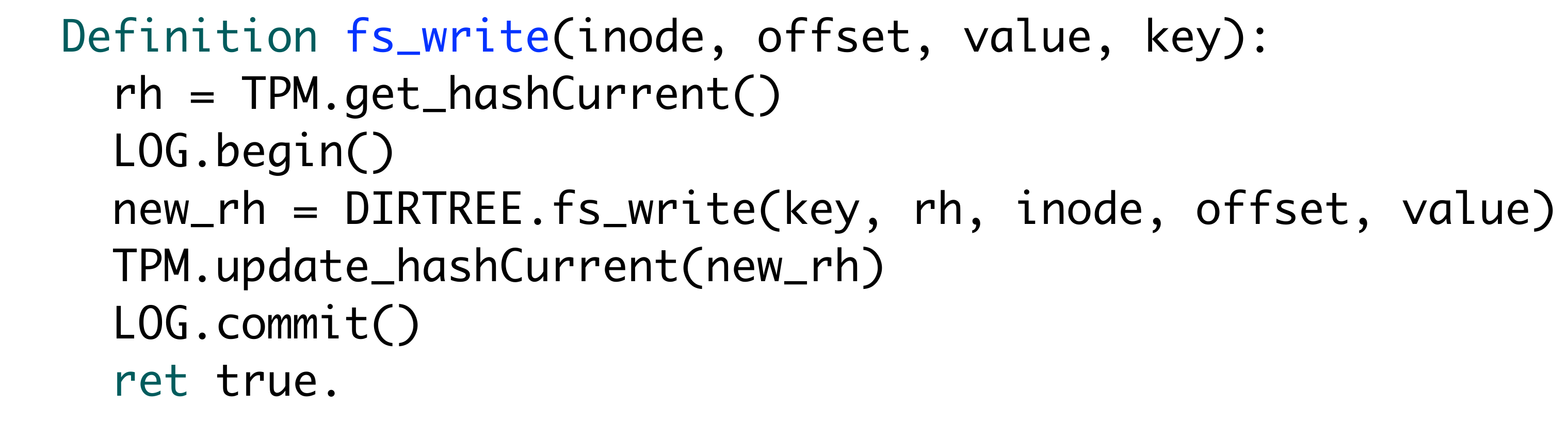}
	\caption{Top-level file write operation.\label{fig:fs_write_code}}
\end{figure}

\vspace{1mm} 
\noindent\textbf{TPM.} The TPM holds three hashes: \texttt{hashSB}, \texttt{hashCurrent}, and \texttt{hashRecover}. \texttt{hashSB} is the hash of the file system superblock; \texttt{hashCurrent} and \texttt{hashRecover} are used to store the root of the Merkle tree. 
We also store an HMAC key in the TPM, which is otherwise only stored in memory. This ensures that an attacker can never
compute a valid HMAC on a new block, but can only reuse old HMACs.

When IFSCQ restarts, it first reads the superblock which holds the metadata of the file system. The hash of the superblock gets compared with the \texttt{hashSB} to ensure there are no inconsistencies during the restart process.  The other two root hashes for the Merkle tree represent the root hashes for the most recent disk image as well as one operation beforehand, corresponding to the most recent transaction in progress as well as the prior stable state of the disk. Consequently, whether or not the regular FSCQ recovery mechanism rolls back the disk state, it will verify correctly with one of the two root hashes.

Figure~\ref{fig:fs_write_code} shows an example where we update the new root hash to the TPM using \texttt{TPM.update\_hashCurrent} just before the commit which will first move the current \texttt{hashCurrent} to \texttt{hashRecover} and update new hash to the \texttt{hashCurrent}.  The \texttt{LOG.commit} can have two outcomes if there is a crash. The recovery could either roll back the disk state to the state before the \texttt{LOG.begin} or it could finish the updates done between the \texttt{LOG.being} and \texttt{LOG.commit}.

\begin{figure}
\centerfloat
	\subfloat[Pseudocode for TPM recovery\label{fig:tpmrecover}]{{\includegraphics[width=8cm]{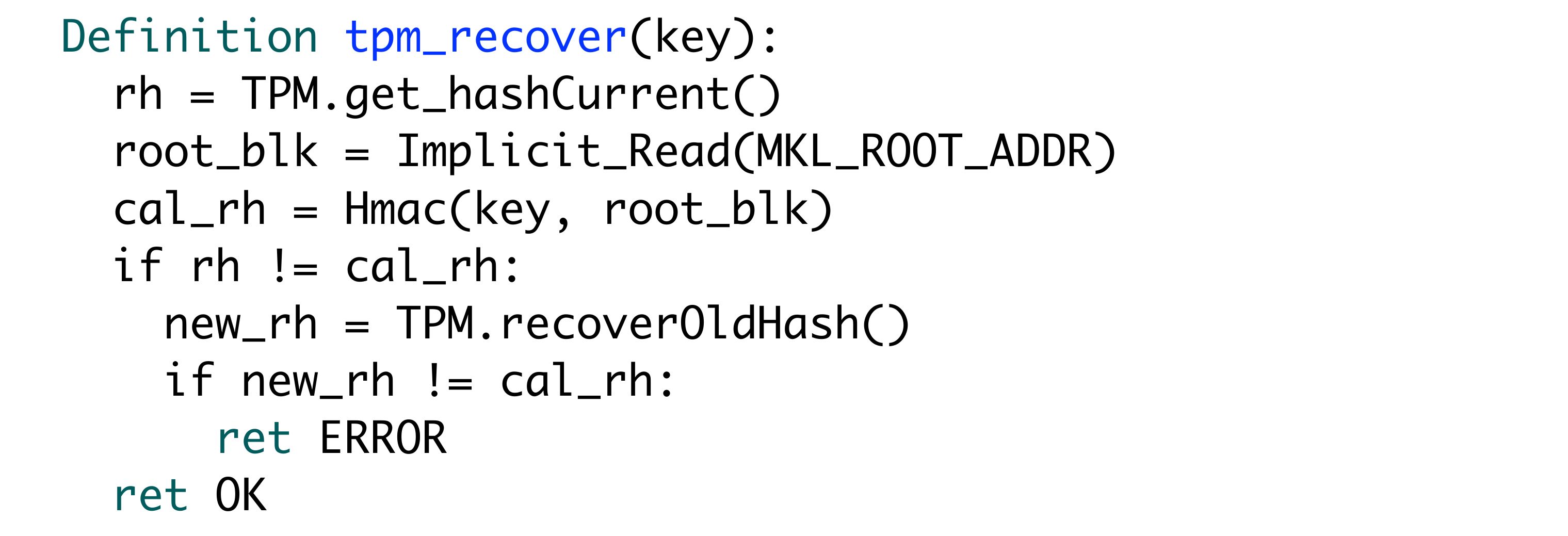} }}%
    \subfloat[SLog read and write example\label{fig:slog_code}]{{\includegraphics[width=8cm]{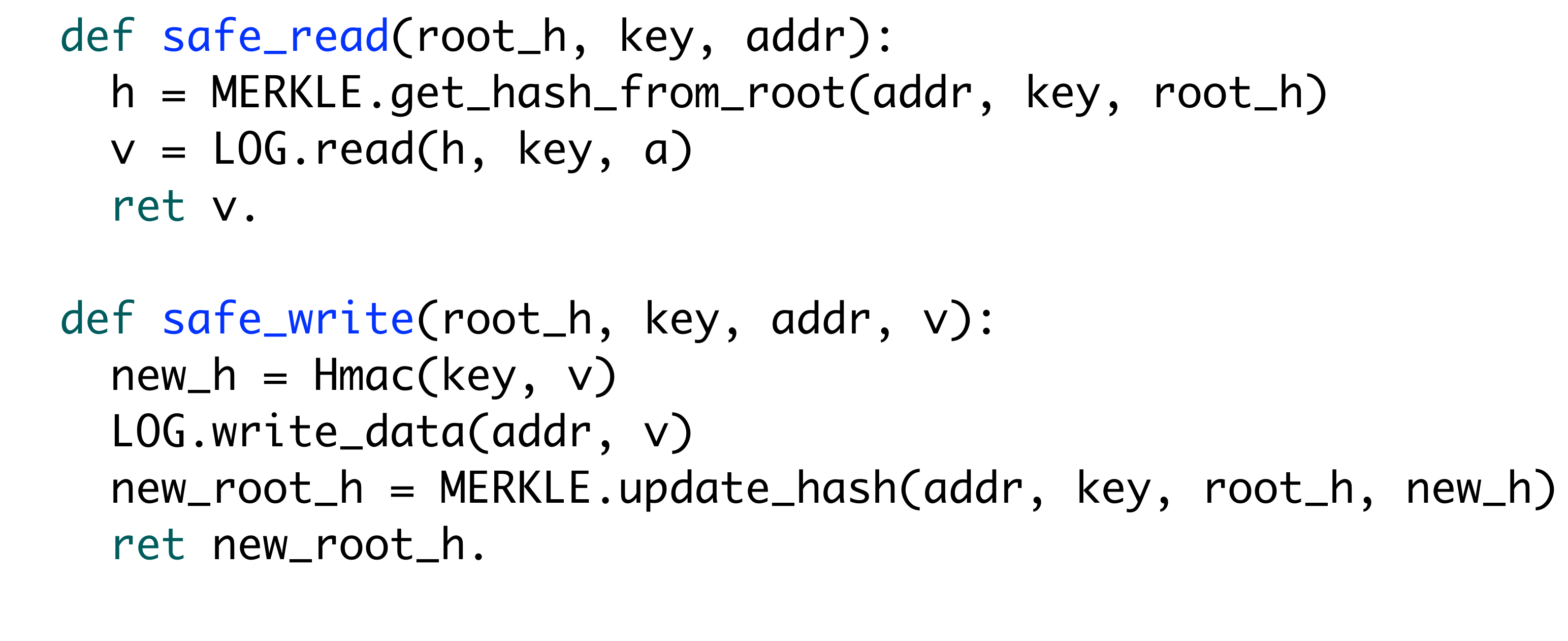} }}%
    \caption{Pseudocode for TPM recovery and SLog read and write\label{fig:tpmandslog}}%
\end{figure}

So what happens if there's a disk crash?
After the regular FSCQ recovery of the disk is done, we call \texttt{tpm\_recover} as shown in Figure~\ref{fig:tpmrecover}, to verify that the root of the Merkle tree, stored on disk, matches with one or the other root hash stored in the TPM. Note that \texttt{TPM.recoverOldHash} has the side-effect of overwriting the ``current'' hash with the ``old'' hash, and if that still doesn't verify, then the recovery operation fails. 

We make an important assumption that the malicious disk is well-behaved during FSCQ's recovery process. Among other things, this means that we're assuming the root block of the Merkle tree, read by \texttt{tpm\_recover}, is correct. Is this a safe assumption? The malicious disk could immediately resume its malicious ways after recovery was complete, so there's no particular incentive for an attacker to break the recovery process, particularly when the attacker has no access to the root hashes stored in the TPM.
(And, on the pragmatic side, it's immensely helpful to avoid worrying about the malicious disk during the recovery process itself.)

\vspace{1mm} 

\noindent\textbf{FS.} The root hash of the Merkle tree, on disk, and the most recent root hash \texttt{hashCurrent} in the TPM should be the same before every top-level procedure call and should also be the same after any updates necessary for disk write operations. The top-level procedures in FS start with \texttt{LOG.begin} and end with \texttt{LOG.commit}, as shown in Figure~\ref{fig:fs_write_code}, guaranteeing that either all of the updates in the transaction get applied or none do. The update operations eventually calls \texttt{LOG.write} which writes a value to the in-memory log instead to the disk. If the in-memory log updates were to be committed without the appropriate updates to the Merkle tree, the disk would have a broken Merkle tree. Thus, we also update the Merkle tree accordingly and retrieve the new root hash (see ``SLog'' discussion, below). When we update the Merkle tree, we must update the TPM. Absent any crashes, the TPM's \texttt{hashCurrent} and the root hash from the Merkle tree disk state should be the same. With a crash but without any malicious disk behavior, one of the two TPM hashes should still validate correctly. With malicious disk behavior, the system halts upon a hash mismatch.




\vspace{1mm} 
\noindent\textbf{SLog, Merkle, Log, and Read.} When we are ``above the Log'', we don't have to worry as much about crash behavior. This is where we put all the checking and validation of our Merkle tree.

\vspace{1mm} 
\noindent\textbf{SLog.} SLog is a wrapper for the Log which provides block read and write operations, calling the Merkle module to read or update the Merkle tree, and ultimately calling the real Log for real disk operations. Components of FSCQ that previously interacted directly with the Log now instead interact with SLog.

The read operation \texttt{safe\_read} in Figure~\ref{fig:slog_code} first retrieves the corresponding leaf hash value using the root hash \texttt{root\_h} of the Merkle tree calling \texttt{get\_hash\_from\_root} which does a top-down approach which checks the nodes along the way, all of which will be checked against their parent's hash. The retrieved leaf hash will then be passed to the log using \texttt{LOG.read} along with the key for the HMAC and the address to read which will eventually call the \texttt{Read} primitive function. The \texttt{Read} primitive function only returns the value when the hash value of the block we are reading matches with the provided hash. For the write, we first calculate the hash of the value we are writing, then write the value to the data block, and finally update the relevant Merkle tree nodes.


\vspace{1mm} 
\noindent\textbf{Merkle and HashRec.} 
Our Merkle tree is a ``full'' or ``complete'' tree. Because we never insert or delete entries, due to the disk's fixed size, the tree is always balanced. We also make our Merkle tree ``fat''; every block holds exactly 128 hashes of child blocks. Furthermore, we can deterministically compute the location of any given leaf in the tree without needing to follow tree pointers, in much the same way that binary heaps (see, e.g., Cormen et al.~\cite{clrs-binary-heap}) can be navigated using index arithmetic rather than pointers.

These properties allow our Merkle trees to be quite compact, minimizing the number of block reads necessary to verify or update a hash. We generally traverse our Merkle trees from the top, downward, except for updates, which we compute starting at the leaf and working our way back up to the root node. One limitation of our current implementation is that we have a fixed-size depth of the tree (currently four), which can handle at most a one terabyte file system. We plan to extend our work to provide variable-depth Merkle trees in the future.

\vspace{1mm} 
\noindent\textbf{Log.} We introduce two write operations to the Log: \texttt{write\_data} and \texttt{write\_hash}. They both update the disk but \texttt{write\_data} is used to update the data part of the disk and \texttt{write\_hash} is used to update the Merkle tree part of the disk. This separation allows us to reason about which part of the disk is being updated. For example, Figure~\ref{fig:safe_write_spec} shows the specification of the \texttt{safe\_write}. The disk of the precondition \texttt{old\_state} is divided into \texttt{old\_data\_disk} and \texttt{old\_hash\_disk} where \texttt{write\_data} will only update the \texttt{old\_data\_disk} and not the \texttt{old\_hash\_disk}. Likewise, \texttt{write\_hash} will only update the \texttt{old\_hash\_disk} and not the \texttt{old\_data\_disk}. This separation simplifies our proofs.

\texttt{LOG.read} eventually reads from the disk if it's not in the cache using a low-level \texttt{Read} primitive. We changed the implementation and specification of the \texttt{Read} primitive so that it requires a hash value for the block we are reading. The \texttt{Read} primitive first reads from a block from the evil disk and calculates the hash. This hash gets compared with the hash that was passed down as an argument of \texttt{Read}, and if it's a match it returns the value or if it's not the system halts.





\section{Evaluation}
\label{sec:evaluation}

In this section, we report results obtained via a set of experiments that are designed to evaluate 
a) the performance of IFSCQ compared to FSCQ, b) the effectiveness in detecting attacks, c) the development effort, 
and d) the correctness of the specification. 
\begin{figure*}
	\center
	\includegraphics[width=\textwidth]{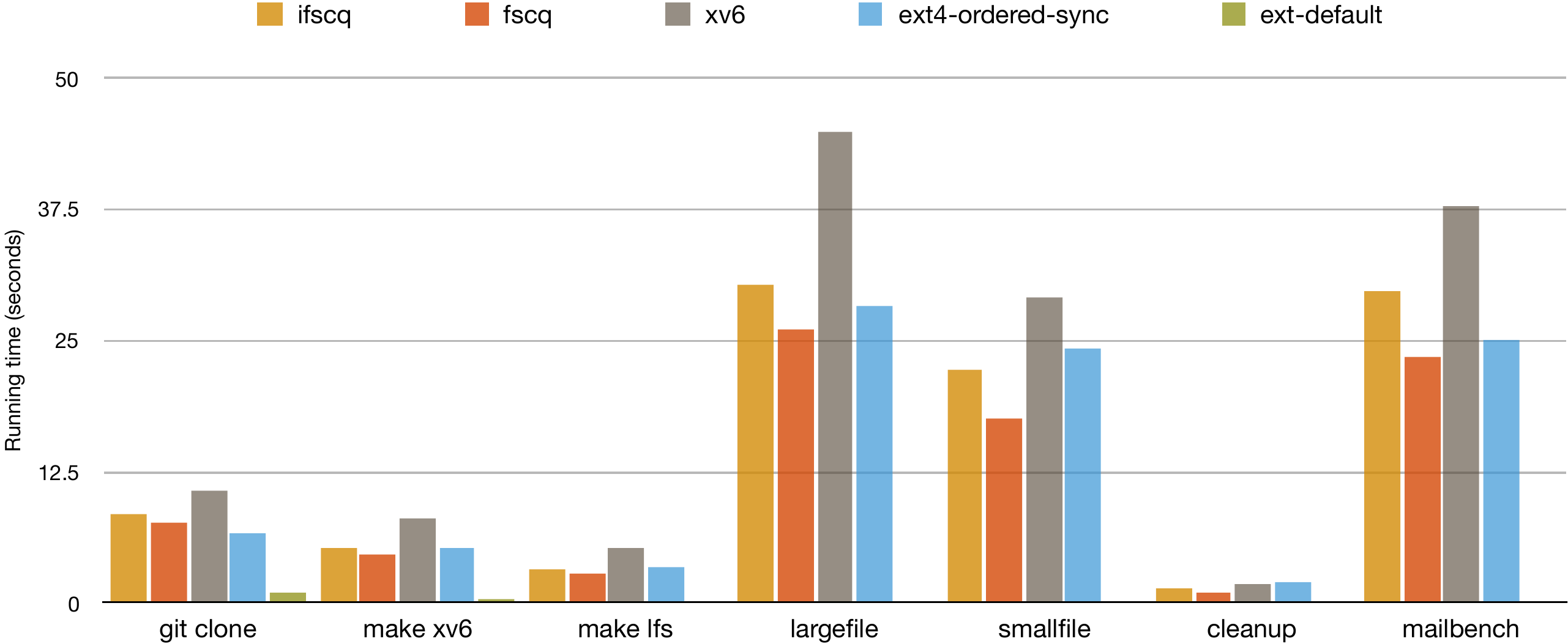}
	\caption{Microbenchmark.\label{fig:benchmark}}
\end{figure*}

\vspace{1mm} 
\noindent\textbf{Performance} We start by evaluating the performance of IFSCQ against several
baseline file systems. While we wish to make it clear that performance
was never a design goal for IFSCQ, we nonetheless provide benchmarks
allowing our results to be compared to FSCQ. It's useful to see the
relative performance, which demonstrates that while our work might not
be fast, it's at least feasible.

While it might be interesting to benchmark IFSCQ against
high-performance systems like ZFS~\cite{zfs} or Btrfs~\cite{btrfs},
this would mostly show that FSCQ doesn't have modern file system
optimizations.

Similarly, while it might be interesting to compare our IFSCQ design
to other possible designs, such as integrating the Merkle tree into
the inodes, we note that building and verifying one version of IFSCQ
was a significant effort. Doing this for multiple designs would be
prohibitive.

\vspace{1mm} 
\noindent\textbf{Performance Overhead.}
We compare the relative performance with other file systems including
FSCQ, xv6 and ext4, using the same set of microbenchmarks in FSCQ
\cite{fscq}.  These microbenchmarks include: cloning a git repository
that contains the files for the benchmark, compiling xv6 and the LFS
benchmark \cite{lfs_bench}, running the LFS benchmark, and then
cleaning all the files from the benchmarks. In addition, we have also
measured the time it takes to deliver 200 messages with mailbench, a
mail server from sv6 \cite{sv6}.

We ran all the file systems except
ext-default in their synchronous modes, much as FSCQ did in their own
evaluation, to make for a more reasonable comparison. 
Also, all five platforms use FUSE, so they all pay the same FUSE
overhead. ext4 is measured using two different options:
``journal and sync'' versus ``ordered and async'' (the default).
We also note that the original FSCQ authors have
improved FSCQ over time. For example, FSCQ's top-level APIs now accept
multiple blocks to read and write instead of a single block, allowing
FSCQ to perform much better for us than their original publication.

We ran our tests on a four-core Intel Core i7 (Ivy Bridge), running at
3.4~GHz with 32~GB RAM and a 256~GB SSD.
As shown in the Figure~\ref{fig:benchmark}, IFSCQ has a modest
performance slowdown compared to the most recent version of FSCQ.
FSCQ does well compared to xv6 due to the optimization added to handle
multiple blocks at once.

One thing missing from this benchmark is the overhead of the TPM.
No modern TPM hardware satisfies our model, so we 
instead substitute a Haskell implementation
which stores the hashes to a file on a separate disk. 

\newcommand{\numop}{

    \begin{tabular}{ l r r r r r }
         & Read & Write & Sync & Hash \\ \hline
        git clone (FSCQ) & 12 & 3,739 & 1,721 & 0 \\
        git clone (IFSCQ)& 20 & 9,361 & 1,721 & 9,817\\ \hline
        make xv6 (FSCQ) & 0 & 1,949 & 1,228 & 0 \\ 
        make xv6 (IFSCQ) & 1 & 5,003  & 1,228 & 3,831\\  \hline
        make lfs (FSCQ) & 0 & 1,121 & 764 & 0 \\
        make lfs (IFSCQ)& 0 & 2,919 & 764 & 1,820\\ \hline
        largefiles (FSCQ) & 2 & 13,299 & 8,196  & 0\\
        largefiles (IFSCQ)& 2 & 34,603 & 8,196 & 25,572\\ \hline
        smallfiles (FSCQ) & 13 & 10,032  & 4,800  & 0\\
        smallfiles (IFSCQ)& 14 & 26,032 & 4,800 & 24,094\\ \hline
        cleanup (FSCQ) & 0 & 546 & 312 & 0\\ 
        cleanup (IFSCQ)& 0 & 1,482 & 312 & 780\\ \hline
        mailbench (FSCQ) & 0 & 18,620 & 8,448 & 0\\ 
        mailbench (IFSCQ)& 0 & 44,594 & 8,368 & 46,780 \\ \hline
    \end{tabular}
    }

\vspace{1mm} 
\noindent\textbf{Write and Hash operation overhead.}
Next, Table~\ref{tab:numop} counts disk operations across our
benchmarks. These consider the number of operations touching the real
disk, rather than any caches. We also present the number of hash
computations (as used for both reads and writes).

The standout result is a factor of 2.5$\times$ in disk writes.
That's half of what we would have expected. We conjecture this is
a result of multiple writes to the same hash blocks being coalesced
into single writes to the disk. So is 2.5$\times$ good or bad?
Given that IFSCQ provides security semantics that FSCQ lacks,
this would seem to be a reasonable price to pay, especially when
performance optimization was only a secondary goal of our work.

\vspace{1mm} 
\noindent\textbf{Storage overhead.}
How much disk space are we ``wasting'' on these hashes? How does this
compare to the overhead we already pay for inodes and other metadata?
We formatted a 128MB disk with both FSCQ and IFSCQ, loading the same
content on each. We found that FSCQ's overhead was 3.1\% while IFSCQ's
overhead was 5.4\%.
This mostly shows that our Merkle tree design imposes a reasonable
cost relative to the overheads we already pay for metadata.

\newcommand{\stoovh}{

    \begin{tabular}{ l r r }
         & FSCQ & IFSCQ\\ \hline
        Storage overhead & 132.02MB (3.1\%) &  135.07MB (5.4\%)  \\
\hline
    \end{tabular}
}

\begin{table}
\centerfloat
\numop
\caption{Number of disk operations for various benchmarks.\label{tab:numop}}
\end{table}

\vspace{1mm} 
\noindent\textbf{Detecting attacks.}  In order to evaluate whether
IFSCQ can indeed detect malicious attacks, we have injected several
kinds of faults to our framework, including malicious modifications to
data blocks, inodes, directory structures, metadata, and wrong hashes
from the Merkle tree in the disk. We ran the ``git'' microbenchmark,
unmounted the disk, manually tampered with the disk, then remounted
it, and re-ran the benchmark. IFSCQ detected each attack.





\newcommand{\detattck}{

    \begin{tabular}{ l r}
          &  Tested attacks \\ \hline
        Provide block-level integrity & Yes \\
        Provide metadata integrity & Yes \\
        Provide integrity of the Merkle tree & Yes \\
        Prevent rollback attacks & Yes \\
      \hline
    \end{tabular}
}

\newcommand{\addedtab}{
    \begin{tabular}{ l r r r }
        Component & LoC(Modeling)& LoC(Impl)& LoC(Proof) \\ \hline
        FS & 0 & 70 & 63 \\
        TPM & 41 & 0 & 0 \\
        Merkle & 60 & 137  & 1,005 \\
        HashRec & 15 & 32 & 512 \\ 
        SLog & 0 & 45 & 426 \\
        Log & 0 & 32 & 144\\
        Hash & 25  & 0 & 0\\        
        Execution Model & 110 & 0 & 0 \\
\hline
    \end{tabular}
 }

\vspace{1mm} 
\noindent\textbf{Engineering observations.}
In our first attempt at building IFSCQ, we tried to create a Merkle tree following the file inode
structures. We got this working up to the root of each file, without
also integrating it into the file directory structure. We abandoned
this approach due to the complexity of its implementation.

We initially began with a current version of DFSCQ, having
many performance optimizations, such as log-bypass
writes. Furthermore, these optimizations would have significantly
complicated our Merkle tree and TPM design and corresponding proofs,
because there would be far more than two possible states to which we
might recover after a crash. Instead, we decided to focus our work
on the stable FSCQ code.

We also measured our effort in terms of added lines of
code (see Table~\ref{tab:loc}). We note that our efforts
represent a small fraction of the complexity of the original FSCQ,
both in terms of the size of the code as well as the proof logic.

\begin{table}
\centerfloat
\addedtab{}
    \caption{Added lines of code.\label{tab:loc}}
\vspace{-10mm}
\end{table}

\vspace{1mm} 
\noindent\textbf{Correctness of the specification.}
Our proofs show that our implementation meets our specification.
However, what if the specification is wrong and how can we detect it?
Unfortunately, there is no formal specification of how POSIX file
system calls should behave, much less for the behavior of a whole file
system. Consequently, we pursued a variety of avenues, including
``fuzzing'' our implementation with a malicious disk where we could
dial in a probability for it to return bad results. These experiments
show that our system correctly detected anomalous disk blocks and
halted. Nonetheless, we have no doubt that correctness, with respect
to our spec, and correctness with respect to the semantics expected
by the universe of Unix utilities, are not one and the same.





\section{Related Work}
\label{sec:relatedwork}

\vspace{1mm} 

\noindent\textbf{Formal verification tools} Two types of verification tools are typically used in building formally verified systems: automated reasoning tools using satisfiability modulo theories (SMT) solvers, and interactive theorem proving tools such as Coq and Isabelle/HOL\cite{isabelle}. 

Interactive theorem proving tools require human assistants to construct proofs as well as the implementations and the specifications. This is a labor-intensive process where the proofs are usually much larger than the implementation itself. Previous verified systems such as seL4~\cite{sel4} required 200~k lines of Isabelle code to prove the 35~k lines of implementation. FSCQ~\cite{fscq} needed a proof that is 10$\times$ larger than its implementation.

For contrast, automated reasoning tools leveraging SMT solvers do not require developers to write proofs. However, the developer must still build a model that is amenable to SMT solving; and because this method requires exploring every possible execution path, the developers need to carefully build the specification and the implementation so that it does not encounter a ``path explosion'' problem wherein the solver cannot complete its computation.

The debate between what tool to choose to verify a system has been going on for a while (see, e.g.,~\cite{model_vs_theorem}), but we note that both techniques have now been used in the creation of verified file systems.

\vspace{1mm} 
\noindent\textbf{Formally-verified systems} Recent advances in formal verification have opened doors to building fully verified systems such as operating systems, compilers, and file systems. seL4~\cite{sel4} is the first formally verified microkernel using Isabelle/HOL. They use refinement~\cite{refinement} proof techniques to prove that the Hoare logic properties of the high-level abstract specification also hold for the low-level executable specification. CompCert~\cite{compcert} is a formally verified compiler built in Coq that compiles a subset of the C programming language. It uses ``simulation relations'' to prove that safety properties proved on the source code hold for the executable compiled code. 

There also have been studies to validate the claims of these verified systems. One study~\cite{compiler_bugs} used fuzz testing to find bugs on different compilers, some verified and others not. They were able to find more than 300 bugs in unverified compilers, such as GCC and LLVM, while none were found on the proven parts of CompCert. Another study~\cite{correct_dis_sys} analyzed three state-of-the-art verified distributed systems: Ironfleet~\cite{ironfleet}, Verdi~\cite{verdi}, and Chapar~\cite{chapar}. 
They tested the implementation with fuzzing tools and also manually inspected the code base to find bugs. They were indeed able to find bugs but the bugs were only present in the unverified parts of each system, mostly at the boundaries between the proven and unproven layers. Perhaps unsurprisingly, the developers had made incorrect assumptions about the code for which they had neither created formal models nor proofs.

\vspace{1mm} 
\noindent\textbf{Formally-verified file systems} Two recent file systems---Yggdrasil~\cite{yggdrasil} and FSCQ/DFSCQ/SFSCQ~\cite{fscq,dfscq,disk_sec} ---both aim to achieve the same goal: building a simple, verified, and crash-consistent file system. The fundamental difference between FSCQ and Yggdrasil lies on the choice of the verification tool and proof methodologies. FSCQ uses Coq, an interactive theorem prover which requires developers to write proofs. 

Yggdrasil uses Z3~\cite{z3}, a popular SMT solver. Yggdrsil notably uses several methods to avoid the path explosion problem. First, Yggdrasil introduces ``crash refinement'' which verifies the set of possible disk states produced by an implementation to be the subset of possible disk states produced by a specification. With this, Yggdrasil does not care how a state has been reached, as long as it is a valid state generated from the specification. 
 Second, the developer has to restrict the implementation 
 to limit the number of states generated. This is also not a trivial process: as reported, the authors spent four months working to come up with a design they could verify, three months to build the file system, and six months to experiment with optimizations. 

The Cogent~\cite{cogent} file system tries to bridge the gap between a verifiable formal model and the low-level executable code. FSCQ, for contrast, has Coq code which cannot be directly executed. Instead, Haskell code is ``extracted'' and runs through a FUSE interface, creating a sizable TCB.  With Cogent, the code describes not only the proof for the high-level formal model but also the translation to the low-level code, resulting in a significant reduction in the size of the trusted computing base.


\vspace{1mm} 
\noindent\textbf{File system with data integrity} The NonStop server~\cite{fault_tolerance} was one of the first systems to use checksums to ensure integrity. The checksums were generated and saved along with the data to be written to the disk. The saved checksum was compared when the data was read back. However, this system did not use a Merkle tree for efficient computing and saved the checksum in the same unreliable disk. Possibly the first security-focused integrity tool was Tripwire~\cite{tripwire}. The system administrator would generate a database of checksums of system critical files and store that database on read-only media (e.g., a CD-ROM). The system could then check this database against the file system as it was running. 
IFS \cite{ifs} is a file system with integrity checking built-in. Unlike our system that only stores the root hash in the persistent storage, this system uses a separate persistent database on top of the file system to store the checksums of the disk blocks. They also do not modify the underlying file system and only provide a layer between the file system and the user to provide integrity. 
Iris~\cite{iris} is a cloud file system that provides data integrity by building Merkle trees following the directory structure. 
Sirius~\cite{sirius} provides data integrity to file systems that are layered over insecure networks.
IPFS~\cite{ipfs} is a peer-to-peer distributed file system that uses Merkle trees.

A variety of commercial file systems use cryptographic hashing with the intent of reliably detecting rare disk failures, as opposed to defending against a malicious storage device. This includes GFS~\cite{gfs}, a replicated file system that can respond to failures by fetching a replica of the data. Similarly, ZFS~\cite{zfs} uses cryptographic checksums to detect disk failures, such as may occur in RAID arrays, to ensure that corrupt data can be detected and, if mirroring is in place, the correct data can be retrieved.


\vspace{1mm} 
\noindent\textbf{Cryptographic file systems} There are a variety of file systems that have focused on using cryptography to encrypt files for data privacy.
The Cryptographic File System (CFS)~\cite{crypto_unix} is one of the first file systems that pushed the encryption features into the file system, rather than requiring the user to run a distinct encryption/decryption tool. This integration saved the user from having to manage the cryptographic key material. Where CFS ran in user-space, and had performance issues, 
NCrypt~\cite{ncrypt} ran in-kernel, for improved performance. Many other file systems have integrated encryption features, including
TCFS~\cite{tcfs} and SNAD~\cite{snad}. The users of these file systems do not see any difference between accessing an encrypted disk and an unencrypted disk.

Two commercial cryptographic file systems that are widely used are Apple's FileVault~\cite{filevault} and Microsoft's BitLocker~\cite{bitlocker}, both of which encrypt the full disk, rather than having individual files encrypted or plaintext. The disk encryption key is typically derived from the user password. Current Apple iOS devices use a dedicated hardware chip to perform the encryption on the bus between the CPU and the Flash chip. An important benefit of these systems is that the file system doesn't need to know anything at all about the underlying encryption scheme.

\section{Discussion}
\label{sec:discussion}

\vspace{1mm} 
\noindent\textbf{Minimizing the TCB.} 
Our work depends on the Coq proof system, Haskell compiler and runtime, and Linux FUSE interface which are all assumed to be ``trusted''. One could argue that our trusted computing base is too large. However, our goal was never to minimize the TCB but to verify the design of our file system. 
There are efforts to minimize the TCB by either proving the whole system stack or by generating code that does not depend on runtime systems at all. 
The DeepSpec~\cite{deepspec} project aims to verify the whole stack: the OS, the compiler and programming language, and even the CPU instruction set.
On the other hand, seL4~\cite{sel4} and Cogent~\cite{cogent} generate C code and prove that it satisfies the Isabelle/HOL specification.
Although our work is focused on verifying the correct design of the cryptographic file system, a great future research challenge would be to minimize the TCB, or port our design to work in one of these other environments.

\vspace{1mm} 
\noindent\textbf{Adding data integrity to the newer FSCQ variants.} 
We initially started this project when DFSCQ~\cite{dfscq} was in an active development. This led to slow progress and we decided to abandon
this for the then-complete FSCQ as our baseline system.
Now that the development of DFSCQ and SFSCQ~\cite{disk_sec} are complete, we could adopt our data integrity feature to these systems.

DFSCQ introduces several optimizations to achieve high performance. Unlike FSCQ, which flushes the log data to the disk when it is committed, DFSCQ buffers transactions in memory, deferring disk writes for increased performance.  These transactions are modeled using a notion of ``tree sequences''---a sequence of file system trees that describe every possible state after a crash.  In order to extend our work to DFSCQ, each of these trees in the sequence would need to correspond to a root hash that we stored in the TPM. Furthermore, if an attacker has access to a long list of ``valid'' restore points, that might yield important new vulnerabilities that would require additional analysis. For example, if a trusted application performed a series of writes which should have been transactional but were not, this would create an opening for a malicious disk adversary to crash-restore the application into an undesired state.

DFSCQ also introduces ``log-bypass writes'' which go immediately to the disk.
DFSCQ models these log-bypass writes by updating every tree sequence in the model that has yet to be flushed to the disk. 
Thus, if we were extend our work to DFSCQ, we not only should maintain the root hash for each tree of the tree sequence but also all of the possible updates to those Merkle trees. Again, this would give additional flexibility to an attacker, as well as creating greater complexity to prove an implementation correct.

SFSCQ is another FSCQ variant that provides confidentiality access controls to a file system using the notion of data noninterference. It introduces the notion of ``sealed blocks'' where it proves that file system to does not read those blocks when not permitted. This would be more straightforward to integrate with IFSCQ, since IFSCQ's integrity checks are largely orthogonal to SFSCQ's confidentiality mechanism.

\vspace{1mm}
\noindent\textbf{Alternate designs and semantics.}
The design space of possible ways of adding integrity checking to file systems is enormous. As we've hopefully already motivated, our goal was to focus on making mostly orthogonal changes to an existing verified file system. Of course, there are many other approaches that could be imagined, such as relaxing the semantics of the file system. If, for example, files were only ever read from start to finish, without random access reads or writes, then we could simply store one hash per file, much like Tripwire. We could even have a separate implementation strategy specifically for append-only log files, perhaps leveraging the latest cryptographic data structures from the block-chain community.

Similarly, our underlying file system might eventually have more
interesting semantics, such as ZFS's ability to take efficient
snapshots and view the old files from those snapshots. Needless to
say, this would necessitate a redesign of our Merkle tree, since we would want to efficiently match up old ``versions'' of the Merkle tree with the corresponding file system snapshots, and we would want sublinear costs for taking the snapshots of the Merkle trees. We would need something more complex than our ``fat'' Merkle design to make this work.



\section{Conclusion}
\label{sec:conclusion}

We have presented IFSCQ, a formally verified cryptographic file system that guarantees data 
integrity against strong adversaries. IFSCQ comes with a formal model of a malicious storage 
system, and proves that the implementation meets the specification of not only the crash 
consistency but also the detection of attacks. 
Evaluation results show that IFSCQ runs with reasonable overhead, and that it can detect 
a range of attacks to the storage system.

%
%
%
\bibliographystyle{splncs04}
%

\bibliography{bib}

\end{document}